\documentclass[useAMS,usenatbib]{mn2e}

\usepackage{amssymb}
\usepackage{graphicx}
\usepackage{txfonts}
\usepackage{color}
\usepackage{amsbsy}

\newcommand{\integral}{{\textit{INTEGRAL}}}

\newcommand{\xmm}{{\textit{XMM}}}

\newcommand{\fermi}{{\textit{Fermi}}}

\newcommand{\lsi}{LS~I~+61\degr 303}
\newcommand{\gr}{$\gamma$-ray}
\newcommand{\msun}{{{\rm M}_{\sun}}}

\newcommand{\ee}{e$^\pm$}

\begin{document}

\title[Pulsar wind nebula model of LS I +61{\degr}303]{A compact pulsar wind nebula model of the $\boldsymbol{\gamma}$-ray loud binary LS~I~+61$\boldsymbol{\degr}$303}

\author[Zdziarski, Neronov and Chernyakova] {Andrzej A. Zdziarski,$^{1}$\thanks{E-mail: aaz@camk.edu.pl (AAZ), Andrii.Neronov@unige.ch (AN), masha@cp.dias.ie (MCh).} Andrii Neronov$^{2,3}$\footnotemark[1] and Maria Chernyakova$^{4}$\thanks{On leave from Astro Space Center of the P. N.~Lebedev Physical Institute, Moscow, Russia.}\footnotemark[1]
  \\$^1$Centrum Astronomiczne im.\ M. Kopernika,
  Bartycka 18, 00-716 Warszawa, Poland \\
 $^2$INTEGRAL Science Data Center, Chemin
  d'\'Ecogia 16, CH-1290 Versoix, Switzerland\\ $^{3}$Geneva
  Observatory, University of Geneva, 51 ch.\ des Maillettes, CH-1290
  Sauverny, Switzerland\\$^4$ Dublin Institute for Advanced Studies,
Fitzwilliam Place 31, Dublin 2, Ireland }

\date{Accepted 2009 December 23.  Received 2009 December 23; in original form 2009 July 15}

\pagerange{\pageref{firstpage}--\pageref{lastpage}} \pubyear{2009}

\maketitle

\label{firstpage}

\begin{abstract} 
We study a model of of \lsi\ in which its radio to TeV emission is due to interaction of a relativistic wind from a young pulsar with the wind from its companion Be star. The detailed structure of the stellar wind plays a critical role in explaining the properties of the system. We assume the fast polar wind is clumpy, which is typical for radiatively-driven winds. The clumpiness, and some plasma instabilities, cause the two winds to mix. The relativistic electrons from the pulsar wind are retained in the moving clumps by inhomogeneities of the magnetic field, which explains the X-ray variability observed on time scales much shorter than the orbital period. We calculate detailed inhomogeneous spectral models reproducing the average broad-band spectrum from radio to TeVs. Given the uncertainties on the magnetic field within the wind and the form of the distribution of relativistic electrons, the X-ray spectrum could be dominated by either Compton or synchrotron emission. The recent {\it Fermi\/} observations constrain the high-energy cut-off in the electron distribution to be at the Lorentz factor of $2\times 10^4$ or $\sim 10^8$ in the former and latter model, respectively. We provide formulae comparing the losses of the relativistic electrons due to Compton, synchrotron and Coulomb processes vs.\ the distance from the Be star. We calculate the optical depth of the wind to free-free absorption, showing that it will suppress most of the radio emission within the orbit, including the pulsed signal of the rotating neutron star. We point out the importance of Compton and Coulomb heating of the stellar wind within and around the $\gamma$-ray emitting region. Then, we find the most likely mechanism explaining the orbital modulation at TeV energies is anisotropy of emission, with relativistic electrons accelerated along the surface of equal ram pressure of the two winds. Pair absorption of the TeV emission suppresses one of the two maxima expected in an orbit. 
\end{abstract}

\section{Introduction}
\label{s:intro}

The Be star binary \lsi\ is one of a few currently known \gr--loud (i.e., with the radiative power dominated by the \gr\ band) X-ray binaries. The spectrum of high-energy emission from the system extends up to TeV energies \citep{albert06} and the peak of its $\epsilon F(\epsilon)$ spectrum, where $F$ is the energy flux and $\epsilon$ is photon energy, is at $\sim\! 100$ MeV.

The origin of the high-energy activity of the source is a subject of a dispute. All binaries known to be accretion-powered have either no detectable emission at $\epsilon\ga 1$ MeV or the \gr\ luminosity much lower than the X-ray one. The latter is the case for Cyg X-1, where weak \gr\ emission was detected in the soft state up to $\epsilon\la 10$ MeV \citep{mcconnell02}, and a weak TeV transient was once detected by the MAGIC detector \citep{albert07}. Also, the \gr\ emission from Cyg X-3 detected in the 0.1--100 GeV range by \fermi\/ \citep{abdo09b} and {\it AGILE\/} \citep{tavani09} is much weaker than the X-ray emission of \lsi. 

It is possible that a \gr\ loudness of an X-ray binary is related to its special orientation to the observer (by analogy with the \gr\ loudness of active galactic nuclei, \citealt{mirabel99}). On the other hand, it is possible that the binaries detected in the TeV \gr\ band are fundamentally different from the accretion-powered X-ray binaries. In fact, one of the \gr\ loud binaries found so far, PSR B1259--63, is known to be powered by the rotation energy of a young pulsar rather than by accretion \citep{johnston92}. In two other \gr--loud binaries, LS 5039 and \lsi, the pulsed emission from the pulsar has not been detected, so there is no direct evidence for the pulsar powering the activity of these sources. However, the binary orbits of these two sources are much more compact than that of PSR B1259--63. Then, the pulsed radio emission would be free-free absorbed in wind from companion star. 

If the activity of \lsi\ is indeed powered by a young pulsar, the radio-to-\gr\ emission is generated in the course of collision of the relativistic pulsar wind with the wind from the companion. Interaction of these winds leads to formation of a scaled-down analogue of the pulsar wind nebulae (PWN) in which the energy of the pulsar wind is released at an astronomical unit, rather than on the parsec, distance scale \citep{maraschi81,tavani97,harrison00,sierpowska05,dubus06,neronov07}.

In this paper, we explore the structure of the compact PWN of \lsi\ in the framework of a model with clumpy stellar wind from its companion Be star and assuming mixing of the pulsar and stellar winds. We calculate the effect of different physical processes that determine energy losses of high-energy particles as a function of the distance from the Be star. This allows us to present spectral models reproducing the average observed spectra from radio to $\gamma$-rays. The two main radiative processes are inverse Compton (IC) scattering of stellar photons and synchrotron emission. Coulomb energy loss is an important process for electrons, especially those with energy in the MeV range. Radio emission from within the orbit of the system is strongly free-free absorbed by the stellar wind. The relativistic electrons are carried away by the stellar wind. Overall, our model explains several important aspects of the behaviour of radio, X-ray and \gr\ emission of the system.

\section{Properties of \lsi\ and its wind}
\label{s:basic}

\subsection{The system parameters}
\label{parameters}

The binary period of the system, from radio observations, is $P = 26.4960\pm 0.0028$~d \citep{gregory02}. Other binary parameters bear larger uncertainties (\citealt{hc81,casares05}, hereafter C05, \citealt{grundstrom07}, hereafter G07). Here, we adopt the values given by G07\footnote{After our calculations were completed, \citet{aragona09} have obtained the binary parameters improved, and slightly different with respect to those of G07, namely $e= 0.54\pm 0.03$, the phase of the periastron of $\phi=0.275$, $\omega=41\pm 6\degr$, and the mass function of $0.012\pm 0.002\msun$. These differences have negligible effect on the results presented here.}, the eccentricity of $e= 0.55\pm 0.05$, the phase ($0\leq \phi\leq 1$) of the periastron of $\phi=0.301\pm 0.011$, and the mass of the Be star as $M_*=12.5\pm 2.5\msun$. (For historical reasons, the phase of $\phi=0$ is defined as corresponding to HJD 2,443,366.775, \citealt{gregory02}.) In the framework of our model, the compact object is a neutron star, which mass we assume to be $M_2=1.5\msun$. Then, the Kepler law yields the semi-major axis of $a\simeq 6.3\times 10^{12}$ cm. G07 also give the radius and the effective temperature of the Be star, of the B0 V spectral classification, as $R_*\simeq 4.7\times 10^{11}$ cm and $T_* \simeq 3.0\times 10^4$ K, respectively, yielding the luminosity of $L_*=4\upi R_*^2 \sigma T_*^4 \simeq 1.3\times 10^{38}$ erg s$^{-1}$, where $\sigma$ is the Stefan-Boltzmann constant. We note that $T_*$ is relatively uncertain and some other studies indicate a lower value, e.g., C05 and \citet{waters88} give $2.8\times 10^4$ K and $2.0\times 10^4$ K, respectively.

As we assume $M_2=1.5\msun$, the mass function of the system, $0.011\pm 0.003\msun$ (C05; G07), implies the inclination of $i=59\degr$, satisfying the constraint of $i\la 60\degr$ of C05, and in agreement with \citet{hc81} (who actually favoured higher values of $i$). Then, the separation at the periastron and apastron is $\simeq 2.8\times 10^{12}$ cm and $\simeq 9.7\times 10^{12}$ cm, respectively. The latter approximately equals the periastron distance of PSR B1259--63, and it is about 15 times smaller than the apastron distance of that system (assuming the mass of its Be star of $10\msun$, e.g., \citealt{tavani97}). The distance to \lsi\ is $d=2.0\pm 0.1$ kpc (\citealt*{dhawan06}, hereafter DMR06; see also \citealt{fh91,steele98}).

\begin{figure} 
\centerline{\includegraphics[width=\linewidth]{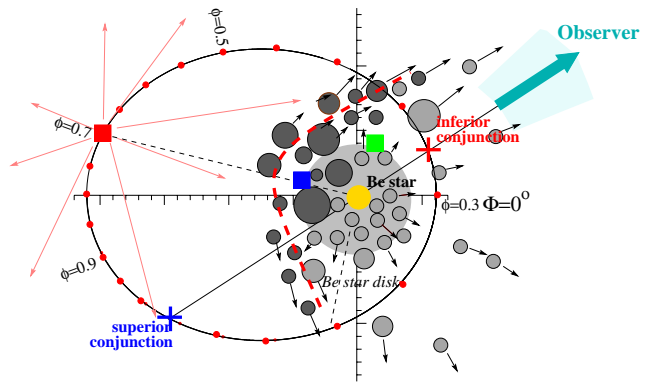}}
\caption{ A schematic representation of the elliptical orbit of \lsi, also showing interaction of the pulsar wind with the clumpy Be-star wind (filled circles). The relativistic pulsar wind hits nearest clumps of the stellar wind. Instead of a smooth bow-shaped contact surface (shown by the dashed curve), the interaction region (shown by the dark circles) is irregularly shaped. The small ticks on the coordinate axes correspond to the radius of the Be star (shown by the yellow circle). The red dots along the orbit mark the phase, $\phi$, spaced with $\delta\phi = 0.05$. The arrow show the direction of the observer, $\phi_{\rm inf}=0.324$, $\Phi_{\rm inf}=33\degr$, the light shaded contour gives its uncertainty of $\pm 9\degr$, and the crosses show the superior and inferior conjunctions. The large grey circle shows the disc of the Be star. The filled red, blue and green squares show the points for which the orbital dependence of the optical depth for $\gamma\gamma$ pair production is calculated, see Section \ref{gamma}.
}
\label{fig:scheme} 
\end{figure}

We also use the true anomaly of the system, $\Phi$ ($0\leq \Phi\leq
360\degr$), which is the angle between the direction from the Be star to
the periastron and that to the compact object. The inclination of the semi-major axis with respect to the line of nodes determined as $\omega=57\pm 9\degr$ (G07) yields the orientation of the observer, and the angular phases of the superior and inferior conjunctions of $\Phi_{\rm sup} = 270\degr - \omega =213\degr$ and $\Phi_{\rm inf}= 90\degr - \omega =33\degr$ (at the best fit), respectively. The corresponding phases are $\phi_{\rm sup}=0.035$ and $\phi_{\rm inf}=0.324$. The components of \lsi\ are shown in Fig.\ \ref{fig:scheme}. 

The mass function of the system together with the estimate of the Be-star mass do not allow us to determine the mass of the compact object, which can still be either a neutron star or a black hole. This is because the inclination of the orbit is only poorly constrained, $10\degr\la i\la 60\degr$ (C05). The compact object would be a neutron star if $i\ga 25\degr$, and a black hole otherwise (C05). In the latter case, the origin of the \lsi\ activity has to be accretion. This type of model was first introduced by \citet{taylor84}, see \citet{boschramon06} for a recent reference. In the former case, it might be accretion as well. However, the source activity may also be due to interactions of a young rotation-powered pulsar with the wind from the companion Be star, as first proposed by \citet{maraschi81}.

The nature of the source has been a subject of an ongoing debate. Main arguments against the accretion scenario are discussed below. Its broad-band spectrum (e.g., \citealt*{cnw06}; \citealt{dubus06}) does not look like that of any black-hole accreting source at a low Eddington ratio (see, e.g., spectra in \citealt{mr06,zg04}). For example, there is no high-energy cut-off in the spectrum at $\sim$100 keV, typical to accreting sources at low
luminosities. Thus, attributing the spectrum of \lsi\ to accretion would require the existence of a yet unknown parameter of accretion, causing black-hole binaries accreting at the same accretion rate to look completely different in TeV binaries and in all other cases. On the other hand, there is the striking similarity between the broad band spectra of all three massive binaries detected as persistent TeV sources, PSR B1259--63, \lsi, and LS 5039 \citep{dubus06}, with the first of the objects seen to contain a non-accreting pulsar. 

We note that \citet{mk09} have argued for a similarity of \lsi\ to accreting black-hole binaries based on the value of the X-ray photon spectral index, which is $\Gamma\sim 1.5$ in the hard state of black-hole binaries as well as it is observed in \lsi. However, a clear difference is the lack of a high-energy cut-off at hard X-rays and its \gr\ loudness in the latter, as discussed above. \citet{mk09} also argued for the presence of a soft, steep power-law X-ray state, common in black-hole binaries, in \lsi\ based on $\Gamma=3.6^{+1.6}_{-1.1}$ (the $1\sigma$ confidence) found in some \integral\/ observations \citep{cnw06}. However, the $2\sigma$ confidence range of those data give $\Gamma$ between 0.7 and 10, as well as no strong blackbody characteristic to that state is present. Contradicting their proposal, \fermi\ finds a high \gr\ flux in all states of the object \citep{abdo09a}, which implies that an extrapolation of the X-ray spectrum up to 0.1 GeV has to have $\Gamma<2$. \citet{mk09} also notice that the index found in the TeV range by the MAGIC, $\Gamma\simeq 2.6$, is similar to the typical X-ray index of black-hole binaries in the soft state. However, this similarity is clearly accidental, and the extrapolation of that spectrum down to the X-ray regime would over-predict the observed X-ray flux by a few orders of magnitude, again contradicting their supposition.

The spectrum of \lsi\ could, in principle, be dominated by the jet emission. However, the system is not viewed close to face-on, so the jet emission is unlikely to fully dominate the spectrum. Furthermore, if there is indeed a jet in the system it is certainly not steady, and it has been claimed to precess fast, changing its orientation on a day time scale \citep{massi04}. Then, there should be periods when the jet is seen away from its axis with the accretion emission dominating the spectrum, but at no time the spectrum has been seen to resemble spectrally a standard accreting black-hole binary.

The changes of the radio spatial structure of the source are correlated with the orbital phase (DMR06). If this is due to jet precession, the precession period would have to be equal to the orbital period, which would be a highly unusual occurrence. In all known cases, the disc/jet precession period in binaries is much longer than the orbital period (e.g., \citealt{l98}). In the case of tidally-induced precession, this is also a theoretical requirement. 

If the compact object were a strongly-magnetized neutron star but {\it accreting}, then the system would have belonged to the class of accreting Be/X-ray binaries (see, e.g., \citealt{coe00,negueruela04} for reviews). However, the spectral properties of \lsi\ are very different from those of accreting X-ray pulsars in Be systems. Although those sources have X-ray spectra below $\sim$10 keV often similar to that of \lsi, they are strongly cut off in hard X-rays (typically with $\sim$20 keV bremsstrahlung-like spectra), with
only very weak emission above $\sim$100 keV (e.g., \citealt{apparao94,kreykenbohm}).

An accreting Be/X-ray binary with the period, eccentricity and spectral type similar to those of \lsi\ is 4U 0115+63/V635 Cas \citep{no01,ziolkowski02}. It contains a B0.2 V star, $P=24.31$ d, $e=0.34$. If the compact object in \lsi\ is a neutron star, then the main difference between the two systems can be the neutron-star spin period, $P_{\rm s}=3.61$ s in 4U 0115+63, which is long enough to permit accretion. Also, its magnetic field (though strong enough not to permit equatorial accretion) is most likely significantly weaker, since it decays during accretion \citep{ug95,kb97}. 

DMR06 have strongly argued for the pulsar-wind nature of \lsi\ based on its VLBA monitoring over one orbital period, showing an extended radio source with the morphology variable within the period and elongated along the axis connecting the stars, as expected in the case of the pulsar wind colliding with the Be stellar wind. (This can also explain the fast radio-structure variability observed by \citealt{massi04}.) The morphology of the source varies much faster at the periastron than at the apastron (but no relativistic velocities have been found), suggesting the direct formation of the radio source by the pulsar wind interacting with the Be stellar wind. In this context, \citet{romero07}, hereafter R07, point out that the the pulsar wind power, $\dot E_{\rm pulsar}\la 10^{31} (B_{\rm s}/10^{12}\,{\rm G})^2 (P_{\rm s}/1\,{\rm s})^{-4}$, where $B_{\rm s}$ is the surface magnetic field, is required to be $\ga 10^{36}$ erg s$^{-1}$ by the observed bolometric luminosity of the system. The corresponding momentum flux is then significantly higher than that of the Be wind, which presents a problem for interpretation of the relatively collimated structures observed by DMR06, as confirmed by the numerical simulations of R07. Indeed, we face the same problem in our calculations below, see Section \ref{gamma}. Thus, this remains to be an unresolved as yet problem of the pulsar-wind model. On the other hand, the numerical simulations by R07 of their black-hole accretion model are unable to explain the shape of the radio structure elongated along the rotating binary axis. 

The results of the accretion simulations in R07 are quite similar to, e.g., those in \citet{okazaki02}, in which case the compact object is a neutron star. In fact, R07 assume both a large radius at which the inner disc is truncated (but they argue this does not affect their results), at which the flow structure would not be affected by the nature of the central compact object, and a low black-hole mass of $2.5\msun$, also not very different from the possible masses of neutron stars. Yet no accreting neutron-star binary has been observed to be similar to \lsi.

R07 also argued that the lack of pulsations detected from \lsi\ represents an argument against the pulsar model. However, as we show in Section \ref{free-free} below, this lack is actually expected given the compactness of the orbit. 

A further evidence for the presence of a young pulsar is provided by a detection of a powerful $\sim$0.2-s flare with the peak 15--150 keV flux of $\simeq 5\times 10^{-8}$ erg cm$^{-2}$ s$^{-1}$ (which is higher by a factor of $\sim 10^3$ than the average source flux) and a blackbody-like spectrum by the {\it Swift}/BAT \citep{dp08,b08}. Such an event is consistent with a soft gamma-ray repeater/anomalous X-ray pulsar burst \citep{dg08}. This would indicate the presence in \lsi\ of a very strong magnetic field, typical to magnetars. Some of those objects do emit pulsar wind, e.g., \citet{gavriil08}, as required in the present case.

An argument for the black-hole nature of the compact object (thus implying accretion as the underlying cause of the source activity) is provided by some measurements of the width of absorption lines in the system. C05 and G07 have found that width to be $v\sin i\simeq 100$--110 km s$^{-1}$. It is most likely due to the rotation of the Be star. Be stars usually rotate at $\sim$0.7--0.8 times the critical rotation velocity \citep{porter96}, which implies $i\simeq 15\degr$--$20\degr$ in the case of \lsi\ though a lower rotation velocity of the star cannot be ruled out (C05). This inclination would then imply the compact object is a black hole. However, G07 argue that those measurements are affected by weak (and time-dependent) emission in the line wings that made the absorption cores appear more narrow. This then can reconcile these results with those of \citet{hc81}, who obtained $v\sin i\simeq 360\pm 25$ km s$^{-1}$, and whose results imply $i\ga 50\degr$ ($M_2\la 2\msun$), in full agreement with the neutron-star nature of the compact object. 

We also note that all Be binaries with known companions contain neutron stars (e.g., \citealt{negueruela04,bz10}). (White dwarf and He subdwarf companions are also predicted, but are very difficult to find.) The reason for that may lie in the system evolution. The rotation, being the defining characteristic of Be stars, may be achieved during a period of Roche-lobe overflow mass transfer from its (initially) more massive companion onto a B star. However, the companion loses most of its mass in the process, becoming a He star with the mass of only a few $\msun$, which, when explodes as a supernova, can leave behind only a neutron star, but not a black hole (see, e.g., \citealt{gies00,tauris06} and references therein). On the other hand, if rotation of Be stars is achieved in a different way, the population synthesis calculations by \citet{bz10} show that Be/black-hole binaries are evolutionary formed only in $\sim\! 1/30$ of cases compared to those with neutron stars, which can explain the observed absence of black-hole systems. Still, the existence of $\la 2$ black holes in known Galactic Be X-ray binaries is possible.

\subsection{Structure of the stellar wind}
\label{wind}

Be stellar outflows have generally two components. One is a radiation-driven and approximately isotropic, fast wind. The other is a slow equatorial outflow forming a thin decretion disc. The process causing the formation of the disc remains unknown \citep{pr03}. The disc seems to be viscous and nearly Keplerian, with the radial velocity much smaller than the rotational one. Observationally, Be decretion discs in binaries have outer radii significantly smaller than those of isolated stars \citep*{reig97,zamanov01}. This appears to be caused by tidal truncation (via Lindblad resonances) of the disc by the neutron star \citep{al94,no01,on01,okazaki02}. Numerical simulations specific to \lsi\ showing tidal truncation of the decretion disc have been performed by R07.

The size of the decretion disc can be measured via the equivalent width of the H$\alpha$ line, $W_\lambda$. In \lsi, $W_\lambda$ varies between $-6$\AA\ and $-18$\AA\ (G07). Using the numerical Be disc model of \citet{gs06} and assuming $i=60\degr$, this yields the disc radius varying between $4R_*$ to $6R_*$. However, C05 found that in order to match the spectrum of the B0 V star in \lsi\ to a template (of the B0 V star $\tau$ Sco), the template needed to be reduced in the flux. If the resulting additional stellar flux comes from the decretion disc (as postulated by C05), $W_\lambda$ needs to be increased (to account for the stellar continuum being lower) by 1.54. This (a relatively uncertain) correction has a minor effect on the disc radius, increasing it to (4.5--$7)R_*$. On the other hand, the separation varies in the (6.1--$21) R_*$ range (Section \ref{s:basic}). Thus, the neutron star can directly interact with the disc only around the periastron, and only when the circumstellar disc reaches its largest size. On the other hand, the disc is not circular, and some disc matter can reach the neutron star around the periastron even if the average radius is less than the separation, as shown in the hydrodynamic simulations of \citet{okazaki02}. 

The density profile of the equatorial disc in \lsi\ has been measured using the IR free-free and free-bound radiation by \citet{waters88} and \citet{mp95}. Assuming the disc half-opening angle of $\theta_0=15\degr$ and $R_*=7\times 10^{11}$ cm, they found the disc ion density of
\begin{equation}
n_{\rm d,i}(D)\simeq n_{\rm d,0} \left(D\over R_*\right)^{-q},
\label{n_disc}
\end{equation}
where $n_{\rm d,0}\sim 10^{13}$ cm$^{-3}$, and $q\simeq 3.2$. Here, $D$ is the distance from the center of the Be star, which we hereafter will take as a characteristic distance. The disc density at the periastron separation ($D\simeq 3\times 10^{12}$ cm) is then $\sim 3\times 10^{10}$ cm$^{-3}$. The actual value of $n_{\rm d,0}$ is relatively uncertain, and it will be somewhat different at our adopted $R_*$, as well as it depends on the unknown value of $\theta_0$. For example, \citet{porter96} found that $\langle \theta_0\rangle$ in Be stars is $5\degr$. Also, $n_{\rm d}$ varies during a long-term cycle of the disc activity. 

Furthermore, it is highly uncertain to what mass-loss rate equation (\ref{n_disc}) corresponds. This depends on the initial outflow velocity, $v_{\rm d,0}$, which has not been measured. \citet{waters88} and \citet{mp95} assumed 5 km s$^{-1}$ and 2--20 km s$^{-1}$, respectively. However, \citet{porter98} found that such outflows would result in a fast removal of the angular momentum from Be stars, and their substantial spin-down, which is inconsistent with rotational velocity distributions for different Be luminosity classes, showing no significant evolution. Thus, likely values of $v_{\rm d,0}$ are $\ll 1$ km s$^{-1}$. For example, \citet{mzw97} used $v_{\rm d,0}\simeq 0.3$ km s$^{-1}$ (and $\theta_0=5\degr$) as the standard disc parameters. At this $v_{\rm d,0}$, the equatorial mass-loss rate is $\dot M_{\rm d}\sim 10^{-8}\,\msun$ yr$^{-1}$, and the outflow velocity at the periastron distance, implied by equation (\ref{n_disc}), is $\sim$3 km s$^{-1}$.

The other outflow component is a radial, radiation-driven, wind. The wind velocity for such outflows follows the usual law,
\begin{equation}
v_{\rm w}(D) = v_{\rm w,0} + (v_{\infty}-v_{\rm w,0})\left(1-{R_* \over D}\right)^{\beta},
\label{v_wind}
\end{equation}
where $v_{\infty}\simeq (1$--$2)\times 10^3$ km s$^{-1}$ and $v_{\rm w,0}\ll v_\infty$ are the initial and terminal, respectively, wind velocity, and $\beta\sim 1$. The number density averaged over possible wind clumping (see below) at $D$ is related to $v_{\rm w}(D)$ and the wind mass-loss rate, $\dot M_{\rm w}$, via the continuity equation, 
\begin{equation}
\langle n_{\rm w,i}(D)\rangle = {\dot M_{\rm w} \over 4 \upi m_{\rm p}\mu_{\rm i} D^2 v_{\rm w}(D)}\simeq {\dot M_{\rm w} \over 4 \upi m_{\rm p}\mu_{\rm i} D (D-R_*)v_\infty},
\label{n_wind}
\end{equation}
where $\mu_{\rm i}\simeq 4/(1+3X)$ is the mean ion molecular weight, which is $\simeq 1.3$ for the H abundance of 0.7, and the second equality is for $\beta=1$ and neglecting $v_{\rm w,0}$. Unfortunately, the wind parameters have not been measured for \lsi. Following the general estimates of \citet{waters88}, we use fiducial values of $\dot M_{\rm w}= 10^{-8}\,\msun$ yr$^{-1}$ and $v_{\infty}= 10^3$ km s$^{-1}$. The kinetic power of the wind is $\dot M_{\rm w} v_{\infty}^2/2\simeq 3\times 10^{33}$ erg s$^{-1}$.

Radiation-driven winds from massive stars are observed to be clumpy, with the clumps filling a fraction of $f\sim 0.1$ of the wind volume, see, e.g., \citet{mr94}, \citet{schild04}, \citet*{bouret05}, \citet{puls06}. Formation of clumps in these winds is predicted theoretically, with hydrodynamical instabilities leading to formation of shocks, density enhancements and rarefactions \citep*{owocki88}. In the case of Be stars, density irregularities on the scale of $\sim\! 10^{11}$ cm were found in the wind of 2S 0114+65 \citep*{apparao91}. The wind density inside the clumps is $1/f$ times that of the smooth wind. For the parameters adopted above and at $D\gg R_*$, we thus have,
\begin{eqnarray}
\label{n_wind2}
n_{\rm w,i} &\simeq& 3\times 10^{8}\left(\frac{v_\infty}{10^8\,\mbox{cm s}^{-1}}\right)^{-1} \left(f\over 0.1\right)^{-1}  \nonumber\\ &\times& 
\left(\frac{\dot M_{\rm w}}{10^{-8}\msun\, \mbox{yr}^{-1}}\right)
\left(\frac{D}{3\times 10^{12}\,\mbox{cm}}\right)^{-2} {\rm cm}^{-3},\\
\langle n_{\rm i}\rangle &\simeq& n_{\rm i} f.
\label{n_wind_av}
\end{eqnarray}
The presence of clumps affects the interaction with the pulsar wind. Also, the rates of two-body processes, e.g., free-free emission and absorption, are enhanced by a factor of $1/f$ with respect to the case of a smooth flow. The detailed form of clumpy winds remains unknown; in particular, it is likely that the space between the clumps is not empty. In that case, the effect of clumping on a process with the rate proportional to the square of a density, $n$, can be described by the generalized clumping factor \citep{oc06}, 
\begin{equation}
f={\langle n\rangle^2\over \langle n^2\rangle},
\label{f}
\end{equation}
where the averages are along a line of sight within a volume of approximately constant properties. If the density is constant within the clumps and the space outside is empty, this $f$ becomes the usual volume filling factor. We note that when individual clumps become optically thick, porosity of the clumpy wind becomes important and reduces the effective opacity \citep{ogs04,oc06,o09}, which complication we neglect in Section \ref{free-free} below.

Apart from the density, the other main parameter of the wind is its temperature, $T$. In the case of a single massive star, $T$ corresponds to radiative equilibrium in the photon field of the star. It is then somewhat lower than $T_*$, and in the case of Be discs, \citet{cb06} find $T\sim 0.6T_*$. For the fast wind, it is likely to be somewhat higher. The wind temperature reaches the Compton temperature of $0.96T_*$ in the full-ionization regime.

However, X-ray and soft $\gamma$-ray emission substantially increases the (optically-thin) wind temperature even if its luminosity is much lower than the stellar one, because the Compton heating rate (in the Thomson regime) is proportional to $\int E F(E)$, where $F(E)$ is the energy flux of a local spectrum. This effect leads, e.g., to an increase of the temperature of a part of the wind close to the X-ray source to values as high as $\sim 10^6$--$10^7$ K in Cyg X-1 and Cyg X-3 \citep{sz07,sz08}. 

We thus calculate the Compton temperature in \lsi. Unlike the case of accreting sources, most of the Compton heating here is done by $\gamma$-rays, not by X-rays. Then, the calculations need to be done employing the full Klein-Nishina cross section rather than using the usual Thomson-limit method (used, e.g., by \citealt*{bms83}). We have calculated the net rate of energy transfer between thermal electrons and a given photon field using the results of \citet{guilbert86}. We use the X-ray/$\gamma$-ray broad-band spectrum as given in Section \ref{s:spectra}, which can be roughly approximated as a doubly-broken power law,
\begin{equation}
{\epsilon F(\epsilon)\over 1\,{\rm erg\, cm^{-2}\, s}^{-1}} =\cases{1.6\times 10^{-13} \left( \epsilon\over 1\,{\rm eV}\right)^{0.47}\!, &$\epsilon\leq 1\,{\rm MeV}$;\cr
1\times 10^{-10} \left( \epsilon\over 1\,{\rm eV}\right)^0, &$1\,{\rm MeV}<\epsilon\leq 10\,{\rm GeV}$;\cr
4.1\times 10^{-3}\left( \epsilon\over 1\,{\rm eV}\right)^{-0.76}\!, &$\epsilon> 10\,{\rm GeV}$\cr}
\label{spectrum}
\end{equation}
(corresponding to the high X-ray state), with the luminosity of $L_{\rm X\gamma}\simeq 6\times 10^{35}$ erg s$^{-1}$ (assuming isotropy and $d=2$ kpc). In addition, there is the stellar blackbody at $T_*=3\times 10^4$ K and $L_*=1\times 10^{38}$ erg s$^{-1}$. We find the Compton temperature for the composite spectrum is $T=4\times 10^5$ K, i.e., more than an order of magnitude higher than $T_*$. Heating by photons with energies $> 1$ MeV dominates that by photons below 1 MeV, in spite of the Klein-Nishina decline of the Compton cross section. 

The above composite spectrum corresponds to points at the same distance from each of the two sources of radiation, and neglects the presumed diffuse nature of the high-energy source. The actual relative proportions of the high-energy spectrum and of the blackbody depend on the position. In a region around the $\gamma$-ray source, the temperature will be substantially higher than the $T$ above. In particular, the Compton temperature for the high-energy spectrum alone, equation (\ref{spectrum}), is as high as $\simeq 7\times 10^8$ K. On the other hand, line cooling will strongly reduce the temperature away from the high energy source (compare the detailed wind structure in Cyg X-1, \citealt{sz07}), which effect increases with the clumping (e.g., \citealt{sz08}). 

The Compton cooling is almost entirely due to stellar photons, and its rate per electron can be written as (e.g., \citealt{bms83}),
\begin{eqnarray}\label{cool_rate}
\lefteqn{C = {k T\over m_{\rm e} c^2} {\sigma_{\rm T} L_*\over \upi D^2}}\\
\lefteqn{\simeq 1.6\times 10^{-16} \left(\frac{T}{4\times 10^{5}\,\mbox{K}}\right) \left(\frac{L_*}{10^{38}\,\mbox{erg\,s}^{-1} }\right) \left(\frac{D}{3\times 10^{12}\,\mbox{cm}}\right)^{-2} {\rm erg\,s}^{-1},\nonumber}
\end{eqnarray}
where $k$ is the Boltzmann constant, $\sigma_{\rm T}$ is the Thomson cross section, and $m_{\rm e}$ is the electron mass. In equilibrium, it equals the heating rate. This can be compared with other relevant rates. Heating of the wind by Coulomb interactions with relativistic electrons (see Section \ref{losses} below) may play a major role here. We find that heating rate, $L_{\rm C}/N$ (where $N$ is the total number of particles in the heated volume, $\sim\! 2 D^3 \langle n_{\rm i}\rangle$), to be comparable to the rate of equation (\ref{cool_rate}) already when the total power supplied to the wind by the Coulomb process is $L_{\rm C}\sim 10^{33}$ erg s$^{-1}$, i.e, a $\sim\! 2\times 10^{-3} L_{\rm X\gamma}$. If $L_{\rm C}\sim 10^{34}$ erg s$^{-1}$, the equilibrium (but neglecting line cooling) wind temperature will be $\sim\! 4\times 10^6$ K. On the other hand, a major cooling effect is due to advection of the thermal energy by the wind away from the central region, $\sim 2 kT v_{\rm w}/D$. At $v_{\rm w}\simeq 10^8$ cm and $4\times 10^5$ K, this rate is $\sim 4\times 10^{-15} (D/3\times 10^{12}\,{\rm cm})^{-1}$ erg s$^{-1}$. These effects will strongly depend on position via the spatial dependence of both the wind density and of the relevant rates. 

These details are beyond the scope of this work, intended to identify the main relevant physical processes. Given the considerations above, we use a fiducial value of the wind temperature of $T=10^5$ K. We note that the measurements of the equatorial mass loss rate discussed above do not depend on $T$ (apart from the Gaunt factor, \citealt{wb75}). On the other hand, free-free absorption strongly depends on $T$, see Section \ref{free-free} below.

\subsection{Free-free absorption}
\label{free-free}

The free-free absorption coefficient due to ions with the atomic
charge, $Z$, is given by (e.g, \citealt{rl79}),
\begin{equation}
\alpha_{\rm ff}= {2^{5/2} \upi^{1/2} e^6\over 3^{3/2} m_{\rm e}^{3/2} c} 
(kT)^{-3/2} Z^2 n_{\rm e} n_{\rm Z} \nu^{-2} \bar g,
\label{aff}
\end{equation}
where $\nu$ is frequency, $n_{\rm Z}$ is the $Z$-ion density, $n_{\rm e}$ is the electron density, $e$ is the electron charge, and $\bar g$ is the average
Gaunt factor. The Gaunt factor for $h\nu\ll kT$ and $\nu \gg \nu_{\rm p}$ equals
\begin{equation}
\bar g ={3^{1/2}\over \upi}\left[ \ln{ (2kT)^{3/2} \over \upi e^2 Z
 m_{\rm e}\nu}-{5\gamma_{\rm E}\over 2}\right],
\label{gaunt}
\end{equation}
where $\nu_{\rm p}$ is the plasma frequency and $\gamma_{\rm E}\simeq
0.5772$ is Euler's constant \citep{spitzer}. Then, using $Z=1$,
$T= 10^5$ K and $\nu=5$ GHz in equation (\ref{gaunt}), and
averaging over $Z$ in equation (\ref{aff}), we find
\begin{equation}
\alpha_{\rm ff} \simeq 0.12 T^{-3/2} {\mu_{\rm i}^2 \over \mu_{\rm e}} 
n_{\rm i}^2 \nu^{-2}\, {\rm cm}^{-1}\simeq 0.175 T^{-3/2} n_{\rm i}^2 \nu^{-2}\, 
{\rm cm}^{-1},
\label{alpha_ff}
\end{equation}
where $T$ and $\nu$ are in units of K and Hz, respectively, and $\mu_{\rm e}=2/(1+X)\simeq 1.2$ is the mean electron molecular weight.

We first calculate the optical depth of the equatorial disc, which density is given by equation (\ref{n_disc}). We consider the optical depth perpendicular to the disc plane. The disc thickness equals $2D\tan \theta_0$. Thus, 
\begin{eqnarray}
\lefteqn{ \tau_{\rm d,ff} \simeq} \\
\lefteqn{\quad 6 \times 10^6 \left(n_{w,0}\over  
10^{13}\,{\rm cm}^{-3}\right)^2 \left(\frac{T}{10^5\,\mbox{K}}\right)^{-3/ 2} \left(\frac{\nu}{1\,\mbox{GHz}}\right)^{-2} \left(\frac{D}{3\times 
10^{12}\,\mbox{cm}}\right)^{-5.4}. \nonumber }
\label{tau_disc_ff}
\end{eqnarray}
Thus, the entire equatorial disc is completely optically thick to radio emission.

Then, we calculate the radial optical depth of the fast wind from infinity down to a given value of $D$. In a clumpy medium, $\tau_{\rm ff}$, is an integral over the square of the density within the clumps times $f$, i.e., $n_{\rm i}^2 f$ (or equivalently, an integral over $\langle n_{\rm i}\rangle^2 f^{-1}$). Using equation (\ref{n_wind2}), we obtain,
\begin{eqnarray}
\tau_{\rm w,ff}&\simeq& 5\times 10^3 \left(\frac{\dot M_{\rm w}}{10^{-8}\msun\, \mbox{yr}^{-1}}\right)^2 \left(\frac{v_\infty}{10^8\,\mbox{cm s}^{-1}}\right)^{-2} \left(f\over 0.1\right)^{-1} \nonumber\\ &\times&
\left( \frac{\nu}{1\,\mbox{GHz}} 
\right)^{-2} \left(\frac{T}{10^5\,\mbox{K}}\right)^{-3/2}
\left(\frac{D}{3\times 10^{12}\,\mbox{cm}}\right)^{-3},
\label{tau_wind_ff}
\end{eqnarray}
which corresponds to $\tau_{\rm w,ff}=1$ at
\begin{eqnarray}
D_{\rm w,ff}&\simeq& 5\times 10^{13} \left(\frac{\dot M_{\rm w}}{10^{-8}\msun\, \mbox{yr}^{-1}}\right)^{2/3} \left(\frac{v_\infty}{10^8\,\mbox{cm s}^{-1}}\right)^{-2/3}  \left(f\over 0.1\right)^{-1/3} \nonumber\\ &\times& 
\left(\frac{\nu}{1\,
\mbox{GHz}}\right)^{-2/3}\left(\frac{T}{10^5\,\mbox{K}}\right)^{-1/2}{\rm cm}.
\label{D_ff}
\end{eqnarray}

Our results imply that (a) the neutron star moves entirely in the optically thick region, explaining the absence of observed radio pulsations and (b) that the radio emission from the system has to be produced at the distances $D>D_{\rm w,ff}$, lying much outside the binary orbit.  The optical thickness of the central region of \lsi\ has been pointed out before, e.g., by \citet{taylor82}, \citet{massi93} and \citet{dubus06}. An additional process absorbing radio photons in the central regions is synchrotron absorption, which is likely to be substantial \citep{leahy04}. On the other hand, the $\sim$10 MeV electrons, responsible for the synchrotron emission from the region of interaction of the pulsar and stellar winds, see Section \ref{losses} below, can escape beyond the distance $D_{\rm ff}$, so that the radio emission from the extended wind interaction region can be much less suppressed than the direct pulsed emission from the neutron star.

\section{Wind interaction}   
\label{interaction}

\subsection{Homogeneous model of the wind interaction}
\label{s:simple}

In the simplest version, the model of interaction of the pulsar wind
with the wind from the companion massive star assumes that an
isotropic relativistic outflow from the pulsar hits a smooth (but
anisotropic, in the case of a Be star) outflow from the star. Geometry
of the interaction surface is bow-shaped, as determined by the
pressure balance between the pulsar and stellar winds. The physical
parameters of pulsar winds remain poorly known. In a simple
model, an isotropic e$^\pm$ wind with a fixed bulk Lorentz factor is assumed. Such a model was developed in details by \cite{tavani97}, and applied to \lsi\ by \cite{dubus06}. 

In this model, the pulsar and stellar winds do not mix, with the boundary between the shocked and un-shocked stellar wind being spatially separated from the boundary between the shocked and un-shocked pulsar wind. This allows the shocked pulsar wind to escape with a speed of $\sim\! c/3$ (which is much higher than the velocity of the shocked stellar wind). The corresponding time scale is,
\begin{equation}
t_{\rm esc}\sim {R_{\rm w}\over c/3} \simeq 300 \left(R_{\rm w}\over 3\times 10^{12}\,\mbox{cm}\right) {\rm s},
\label{escape_smooth}
\end{equation}
where $R_{\rm w}$ is the characteristic smooth wind size (taken as the periastron separation). However, the power of the pulsar wind itself is constant. This means that any variability is determined by the details of the pulsar wind interaction with the stellar wind rather than by the above time scale. The system properties change on the orbital time scale, 
\begin{equation}
\Delta t\sim\frac{R_{\rm w}}{v_{\rm orb}}\sim 300 \left(\frac{R_{\rm w}}{3\times 10^{12}\,\mbox{cm}}\right)\left(\frac{v_{\rm orb}}{10^7\, 
\mbox{cm\,s}^{-1}} \right)^{-1}\mbox{ks,}
\label{orbital_time}
\end{equation}
where $v_{\rm orb}$ is the pulsar orbital speed.

\subsection{Inhomogeneity of the stellar wind and short time-scale variability} 
\label{s:short}

A long \xmm\/ observation of \lsi\ in 2005 has revealed variability of the system at the time scale, $\Delta t$, of several ks \citep{sidoli06}, which is much shorter than the orbital period. Variability at a similar time scale is also observed in the radio band \citep*{peracaula97}. This implies the presence of a variability mechanism much faster than overall changes of the system
properties with the orbital motion, equation (\ref{orbital_time}). This is possible if the characteristic size scale of variability of the stellar wind is much smaller than the binary separation. This implies that the stellar wind is clumpy rather than smooth, as we have argued on independent grounds in Section \ref{wind}. Furthermore, recent results of \citet{smith09} have also shown rare events in the form of flares with the doubling time scale as short as $\ga\! 2$ s. 

If the wind is clumpy, then the observed variability time scale of $\sim\! 10$ ks is likely to correspond to the time of the passage of a clump through an interaction region. Using $v_{\rm w}\sim 10^8$ cm s$^{-1}$ of the polar wind, the smaller of the required clump size, $R_{\rm w}$, and the size of the interaction region is $\sim v_{\rm w}\Delta t\sim 10^{12}(v_{\rm w}/10^8\,\mbox{cm\,s}^{-1}) (\Delta t/10\,\mbox{ks})$ cm. However, the characteristic size of a clump can be smaller if the observed variability results from averaging over passages of a number of the clumps. The $\ga\! 2$-s time scale may correspond to very rare events when the pulsar wind
manages to irradiate a few clumps with a high magnetic field, as expected close to the stellar surface. \citet{smith09} also discuss the possibility of this time scale corresponding to the stand-off distance between the pulsar and stellar winds. 

Furthermore, the clumpy stellar wind and the pulsar wind are very likely to mix, e.g., due to plasma instabilities. Then, magnetic field can retain the high-energy particles within the clumps, as we show in Section \ref{losses} below. Then, their escape velocity slows down to the stellar wind velocity. The corresponding escape time is,
\begin{equation} 
  t_{\rm esc}\sim D/v_{\rm w}\sim 3\times 10^4\left(\frac{D}{3\times 10^{12}\,
\mbox{cm}}\right)\left(\frac{v_{\rm w}}{10^8\,\mbox{cm\,s}^{-1}}\right)^{-1}\,
\mbox{s.} 
\label{tesc}
\end{equation}

It is not clear whether the multi-TeV electrons responsible for the
observed TeV \gr\ emission are initially present in the un-shocked
pulsar wind, or they are produced via shock acceleration in the wind
collision. Another mechanism of injection of relativistic electrons
can be proton-proton collisions \citep{neronov07}. Here, we thus assume an
injection of relativistic electrons and study its consequences in the presence of wind mixing. 

\begin{figure}
\includegraphics[width=\linewidth]{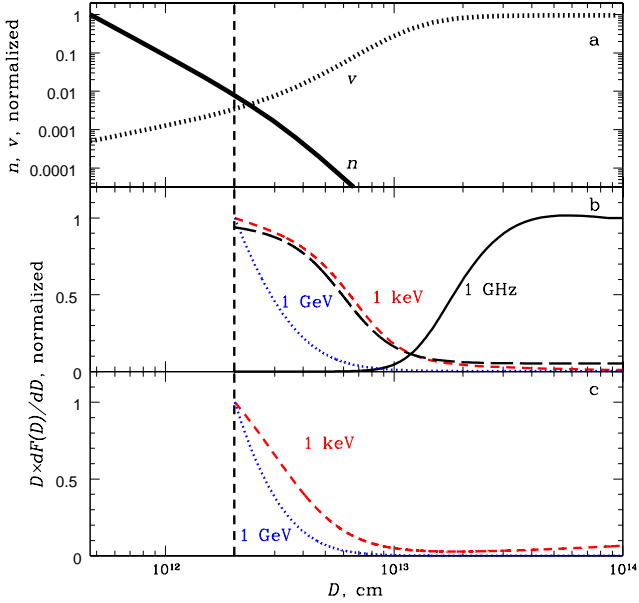}
\caption{(a) The radial profiles of the stellar-wind density and velocity (each normalized to unity at the maximum) assumed in the spectral calculations shown in Fig.\ \ref{fig:LSI_SED1}. The resulting normalized flux radial profiles [per $\ln(D)$] are shown in (b), (c) for the cases computed in Fig.\ \ref{fig:LSI_SED1}(a) and (b), respectively. The vertical dashed line marks $D_0$, the assumed lower boundary of the energy injection region. The black long-dashed curve in (b) shows the fraction of energy of 10-MeV electrons going into Coulomb losses. There is no 1-GHz curve in (c) because that model yields virtually no radio emission. 
}
\label{fig:radial_profile}
\end{figure}

\subsection{Spectral models}
\label{s:spectra}

Below we calculate spectra resulting from superposition of contributions from different radii. Before this, we provide a general characterization of the source in terms of the photon and magnetic compactnesses (e.g., \citealt{vp09,mb09}) and the Thomson optical depths. These compactnesses are,
\begin{eqnarray}
\ell_{*}&=& {L_* \sigma_{\rm T}\over D m_{\rm e} c^3}\simeq 10^{-3} \left( L_*\over 10^{38}\,{\rm erg\, s}^{-1} \right) \left(\frac{D}{3\times 10^{12}\, \mbox{cm}}\right)^{-1},\\
\ell_{{\rm X}\gamma}&=& {L_{{\rm X}\gamma} \sigma_{\rm T}\over D m_{\rm e} c^3}\simeq 10^{-5} \left( L_{{\rm X}\gamma}\over 10^{36}\,{\rm erg\, s}^{-1} \right) \left(\frac{D}{3\times 10^{12}\,\mbox{cm}}\right)^{-1},\\
\ell_{B} &=& {(B^2/8\upi) \sigma_{\rm T} D \over m_{\rm e} c^2}\simeq 10^{-7} \left( B\over 1\,{\rm G} \right)^2 \left(\frac{D}{3\times 10^{12}\, \mbox{cm}}\right),
\end{eqnarray}
where $B$ is the magnetic field strength. Thus, all the compactnesses are $\ll 1$, and $\ell_B\ll \ell_{{\rm X}\gamma}\ll \ell_*\ll 1$. In the Thomson regime, single Compton up-scattering of the stellar photons is usually the main radiative process. However, electrons with the Lorentz factor of $\gamma\ga 10^5$ are in the Klein-Nishina regime, and then synchrotron losses may dominate. The magnetic field is typically below equipartition with the radiation energy density.

The equatorial Thomson optical depth in the disc component of the wind, equation (\ref{n_disc}), from $D$ to infinity (neglecting the disc truncation) is given by,
\begin{equation}
\tau_{\rm d}\simeq 0.02 \left( n_{\rm d,0}\over 10^{13}\,{\rm cm}^{-3}\right)  \left(D\over 3\times 10^{12}\,\mbox{cm} \right)^{-2.2}.
\label{tau_disc}
\end{equation}
The radial Thomson optical depth in the radial component of the wind, equation (\ref{n_wind}), from $D$ to infinity (neglecting its truncation) is even lower,
\begin{equation}
\tau_{\rm w}\simeq 6\times 10^{-5} \left(\frac{v_\infty}{10^8\,\mbox{cm s}^{-1}}\right)^{-1}\! \left(\frac{\dot M_{\rm w}}{10^{-8}\msun\, \mbox{yr}^{-1}}\right) \left(\frac{D}{3\times 10^{12}\,\mbox{cm}}\right)^{-1}\!.
\label{tau_wind}
\end{equation}
Thus, the Thomson optical depths of interest are $\ll 1$. 

\begin{figure*}
\includegraphics[width=0.7\linewidth]{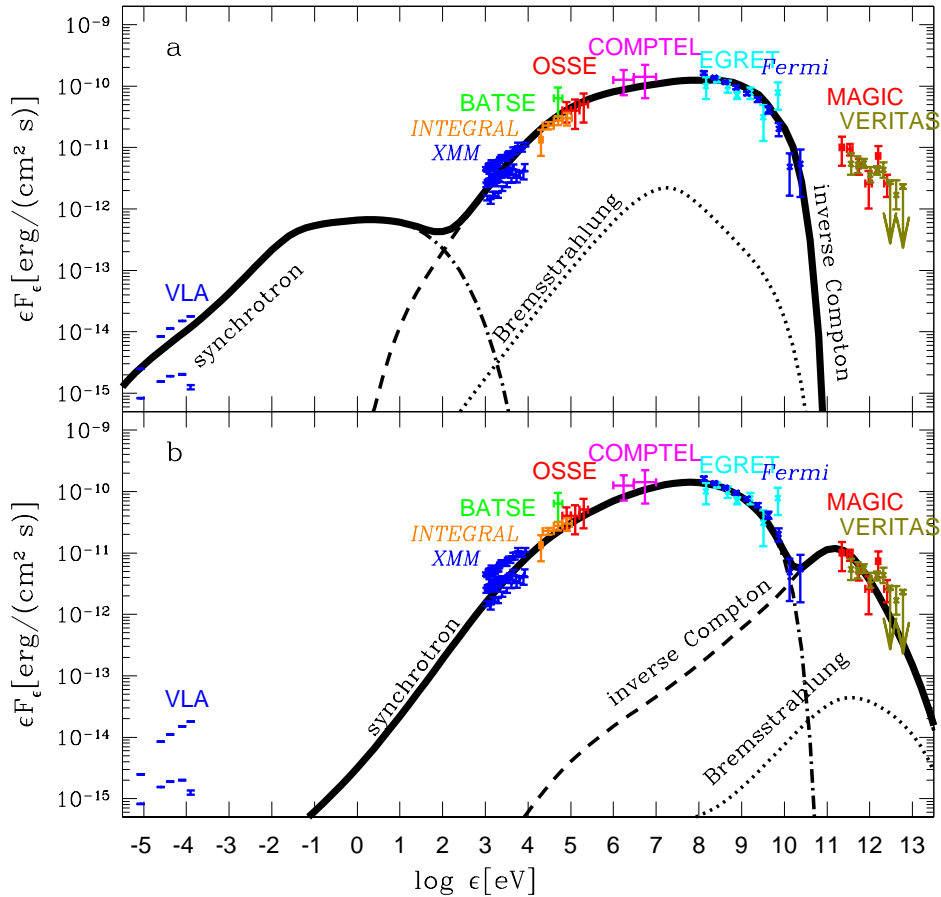}
\caption{Model spectra compared to the data, which are the same as those in \citet{cnw06} except for the data from the VERITAS \citep{veritas} and {\it Fermi\/} \citep{abdo09a} instruments. The dot-dashed, dotted and dashed curves show the spectral components from the synchrotron, bremsstrahlung and IC processes, respectively, and the thick solid curve gives the sum. (a) The model in which the X-rays are dominated by the IC process. (b) The model in which the X-rays are dominated by the synchrotron process. See Section \ref{s:spectra} for the parameters. }
\label{fig:LSI_SED1}
\end{figure*}

We have constructed a one-dimensional spectral model of the source, taking into account the synchrotron and bremsstrahlung emission, free-free absorption by the stellar wind, and Compton, Coulomb, and photon-photon e$^\pm$ pair production processes. We consider a region of the wind in the equator moving with the wind velocity (i.e., assuming that the relativistic electrons are trapped in the wind, see Section \ref{losses} below). Then, the time evolution of the electron spectrum corresponds also to the spatial movement. As our one-dimensional model is taken to represent the entire source, the considered volume corresponds to a conical region centred on the star. We thus use the standard kinetic equation for the evolution of a relativistic electron distribution in the presence of radiative losses \citep{bg70}, in which we replace the time derivative, $\partial/\partial t$, by the spatial one, $v(D)\partial/\partial D$. The escape of photons is also treated one-dimensionally, outward along the radial coordinate.

We follow the evolution of electron spectrum in the distance range of $D_0 = 2\times 10^{12}\,{\rm cm}\le D\le 10^{14}$ cm. The power in the injection per unit radius in the considered conical volume is assumed to be $\propto D^{-1}$. The electron injection spectrum is assumed to be an e-folded power law, ${\rm d}\dot N_{\rm e}/{\rm d}E\propto (E+m_{\rm e}c^2)^{-\Gamma_{\rm e}}\exp(E/E_{\rm c})$ at $E>0$, where $E$ is the electron energy, and $E_{\rm c}$ is its cut-off value, $E_{\rm c}\equiv\gamma_{\rm c} m_{\rm e} c^2$. The power-law injection spectrum may originate in the shock region (e.g., \citealt{ba00}), resulting from the wind collision (e.g., \citealt{k07}). 

The radial profiles of the density and velocity of the wind assumed for the spectral calculations are shown in Fig.\ \ref{fig:radial_profile}(a). The actual wind contains the disc and radial components, see Section \ref{wind}, which would be difficult to take into account in one-dimensional calculations. Therefore, close to the star, at $D\le 6R_*$, we consider only the dense circumstellar disc, with the density following equation (\ref{n_disc}) assuming $n_{\rm d,0}=8\times 10^{12}$ cm$^{-3}$ and $q=3.2$. The initial velocity of $v_{\rm d,0} =10^5$ cm s$^{-1}$ is assumed, which increases $\propto D^{q-2}$. Then, the disc is truncated at $D=6R_*$ and outside it the wind asymptotically reaches the terminal velocity of the radial wind of $v_\infty=10^8$ cm s$^{-1}$. Thus, the density decreases at large distances as $n(D)\propto D^{-2}$, see equation (\ref{n_wind}). For simplicity, we neglect in these models the effect of the wind clumpiness on the radiative rates. 

The magnetic fields of Be stars remain mostly unknown \citep{n07}, though dipole magnetic fields of several hundred G were claimed in some cases (\citealt{st03} and references therein). On the other hand, the upper limits on the Be-star fields found by \citet{hubrig07} are at the level of tens of Gauss. Then, a dipole spatial dependence would imply the field at $D_0$ of $\la 1$--10 G. At $D\geq D_0$, we assume $B=B_0 (D_0/ D)$, and adopt $B_0=2.3$ G.

Fig.\ \ref{fig:LSI_SED1} shows the results of calculations for two choices of the electron injection. Figs.\ \ref{fig:LSI_SED1}(a), (b) show results for the cases of $\Gamma_{\rm e}=1.7$, $E_{\rm c}=10$ GeV ($\gamma_{\rm c}\simeq 2\times 10^4$), and $\Gamma_{\rm e}=1$, $E_{\rm c}=50$ TeV ($\gamma_{\rm c}\simeq 10^8$), respectively. In the two cases, the dominant contribution to the 1 keV--10 GeV spectrum is the IC and synchrotron emission, respectively. Both models fit well the \fermi\/ 0.1--100 GeV spectrum \citep{abdo09a}, in particular, its relatively sharp high-energy cut-off. The bolometric luminosities are $\simeq 5.4\times 10^{35}$ erg s$^{-1}$ and $5.1\times 10^{35}$ erg s$^{-1}$, respectively [isotropic at $d=2$ kpc, and in approximate agreement with the approximation of equation (\ref{spectrum})], and the contribution of the non-thermal bremsstrahlung is minor in both cases. 

In the case shown in Fig.\ \ref{fig:LSI_SED1}(a), the high-energy cut-off in the electron distribution at $\gamma_{\rm c}$ yields a spectral cut-off in the Compton spectrum reproducing the \fermi\/ data at $\epsilon\ga 3 kT_* \gamma_{\rm c}^2\simeq 3$ GeV. An additional minor effect increasing the sharpness of the cut-off is the onset of the Klein-Nishina regime of Compton scattering, at $\epsilon \ga (m_{\rm e}c^2)^2/(3 kT_*) \simeq 30$ GeV or so. We see that the model with the dominant IC emission fails to explain the TeV emission, which then needs to be accounted for by a separate spectral component. On the other hand, preliminary results of observations of \lsi\ by VERITAS contemporaneous with those by \fermi\/ have yielded no detection, with the upper limits on the 1 TeV flux a factor of $\sim\! 2$ below the fluxes from MAGIC and VERITAS at other epochs. Thus, the TeV emission during the \fermi\/ observations was at most weaker than that shown in Fig.\ \ref{fig:LSI_SED1}. 

The steady-state electron distribution, ${\rm d}N_{\rm e}/{\rm d}E$, is determined by the form of the injection and the energy loss and escape processes. Then, the break in the bremsstrahlung spectrum around $\epsilon\sim\! 30$ MeV is due the transition from the dominance of Coulomb losses at $E_{\rm b}\la 30$ MeV ($\gamma_{\rm b}\simeq 60$) to the dominance of Compton losses at higher electron energies. In the former regime, roughly ${\rm d}N_{\rm e}/{\rm d}E \propto {\rm d}\dot{N}_{\rm e}/{\rm d}E$, while in the latter, ${\rm d}N_{\rm e}/{\rm d}E$ is steeper by unity, $\propto \gamma^{-2.7}$. This break is also reflected in the Compton spectrum, at $\epsilon\sim 3k T_* \gamma_{\rm b}^2\simeq 30$ keV.

We see that the synchrotron component of this model accounts quite well for the radio flux. The peak of this spectrum corresponds to characteristic energy of synchrotron photons emitted by an electron with $\gamma_{\rm c}$, $\epsilon\simeq (3/4\upi)\gamma_{\rm c}^2 (h e B_0/2 m_{\rm e} c)\sim 10$ eV, where $h$ is the Planck constant, $e$ is the electron charge, and $1/2$ accounts for the average sine of the pitch angle. The part of the steady-state electron distribution emitting the synchrotron emission is an e-folded power law, and then can be calculated analytically (see eq.\ 
9 in \citealt{z03}). The break at $\epsilon \ga 0.01$ eV ($\nu\ga 2\times 10^{12}$ Hz) corresponds to the onset of free-free absorption by the stellar wind, see Section \ref{free-free}. 

In the case shown in Fig.\ \ref{fig:LSI_SED1}(b), the spectral component dominant at $\epsilon \la 20$ GeV is synchrotron emission. Its dominance over Compton scattering is due to the electron injection being strongly concentrated at $\gamma_{\rm c}\simeq 10^8$, which is very deep in the Klein-Nishina regime (which corresponds to $\gamma\ga 10^5$). This strongly reduces the Compton losses and allows the synchrotron losses to dominate at electron energies responsible for most of the synchrotron emission. The peak of it is close to the characteristic energy of $\epsilon\simeq (3/4\upi)\gamma_{\rm c}^2 (h e B_0/2 m_{\rm e} c)\sim 200$ MeV (see above). The steady-state electron distribution is approximately $\propto\gamma^{-2}$ at both the regimes dominated by synchrotron losses and by Thomson losses ($\gamma\la 10^5$), but at different normalization in each regime, with the regime dominated by Klein-Nishina losses in-between.

Then, the Compton component dominates at photon energies at which synchrotron emission is by electrons deep in the exponential tail of their distribution, $\ga 10\gamma_{\rm c}$, $\epsilon\ga 20$ GeV. The Compton component breaks at the transition from the Thomson to Klein-Nishina scattering, $\epsilon \sim 100$ GeV, corresponding to $\gamma\sim 10^5$. This is also the onset of substantial pair absorption on the stellar photons. On the other hand, the present model cannot account for the radio emission, which needs to be due to a different physical component. Given that the previous model accounts for the radio emission but not for the TeV one, the actual physical conditions in the source may correspond to a superposition of the two models considered here. 

An important issue is the origin of the very hard injection and the high maximum Lorentz factor, $\gamma_{\rm c}\simeq 10^8$ in this model. Very hard, almost mono-energetic acceleration rates have been predicted theoretically and invoked to explain some astrophysical phenomena, see, e.g., \citet{lr92}, \citet{p04}. However, a problem is presented by the high value of $\gamma_{\rm c}$. As noted by \citet*{gfr83}, the acceleration rate is generally $\propto B$, the synchrotron energy loss is $\propto B^2$, and the synchrotron photon energy is $\propto B\gamma^2$. Combining these dependencies lead to the maximum possible characteristic synchrotron photon energy from accelerated electrons being independent of $B$, 
\begin{equation}
E_{\rm max, S}={9\xi\over 32} {m_{\rm e} c^2\over \alpha_{\rm f}}\simeq 20 \xi\,{\rm MeV},
\label{emaxs}
\end{equation}
where $\xi<1$ is an acceleration efficiency and $\alpha_{\rm f}$ is the fine-structure constant (cf., e.g., eq.\ 8 in \citealt*{zmb09}, where, however, no average over the pitch angle was included). On the other hand, the characteristic synchrotron energy in the spectrum in Fig.\ \ref{fig:LSI_SED1}(b) is $\simeq 200$ MeV, 10 times higher even if we take the maximum efficiency of $\xi=1$. Thus, a different mechanism than shock acceleration is needed. This can be due to direct injection of electrons from cold relativistic pulsar wind with the bulk Lorentz factor equal to $\gamma_{\rm c}$. Alternatively, the electrons could be produced in interactions of high-energy protons (e.g., \citealt{neronov07,neronov09}).

\begin{figure}
\centerline{\includegraphics[width=\linewidth]{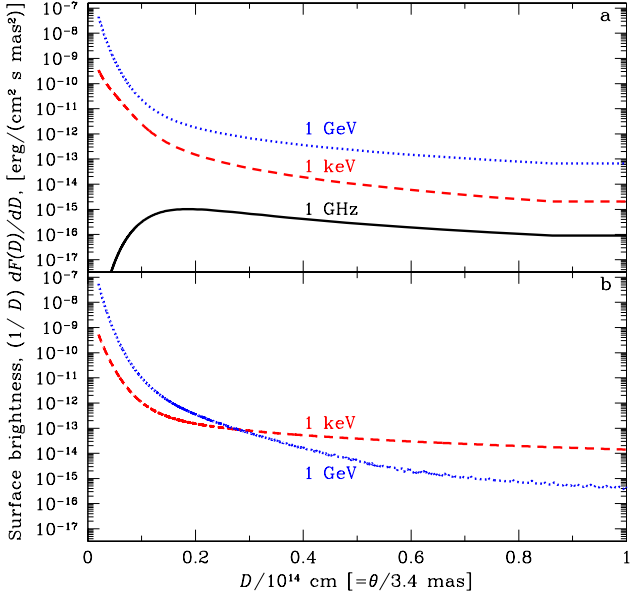}}
\caption{The surface brightness profiles in three energy bands, with (a) and (b) corresponding to the radial profiles and spectra in Figs.\ \ref{fig:radial_profile}(b), \ref{fig:LSI_SED1}(a) and Figs.\ \ref{fig:radial_profile}(c), \ref{fig:LSI_SED1}(b), respectively. There is no 1-GHz curve in (b) because that model yields virtually no radio emission.}
\label{fig:brightness}
\end{figure}

Figs.\ \ref{fig:radial_profile}(b--c) show the normalized radial flux profiles at photon energies in the radio, X-ray and high-energy $\gamma$-ray regimes, corresponding to the models shown in Figs.\ \ref{fig:LSI_SED1}(a--b), respectively. We plot here the flux per logarithm of the distance. We see that whereas the X-ray and $\gamma$-ray fluxes are dominated by the contributions from around $D_0$, the radio flux in the model of Fig.\ \ref{fig:LSI_SED1}(a) peaks above $10^{13}$ cm. We also show the flux profiles in a different representation in Fig.\ \ref{fig:brightness}, in which the actual fluxes are plotted in such units that they are proportional to the surface brightness. 

We have also quantified the role of Coulomb losses of relativistic electrons for the model of Fig.\ \ref{fig:LSI_SED1}(a). First, we have calculated the fractional energy loss going into Coulomb losses for 10-MeV electrons, shown by the black long-dashed curve in Fig.\ \ref{fig:radial_profile}(b). It is $>0.5$ at $D< 6\times 10^{12}$ cm. Then, the corresponding profile of 1 keV emission would be very strongly centrally peaked if Coulomb losses were not included. We have also calculated the spectrum assuming no Coulomb losses (not shown here), in which the effect of Coulomb losses is seen for $1\,{\rm eV}\la \epsilon \la 10\,{\rm MeV}$. That spectrum is approximately flat [in $\epsilon F(\epsilon)$] already at $\epsilon\ga 10$ eV. The power going into Coulomb heating of the wind in this model is very substantial, $\simeq 1.5\times 10^{35}$ erg s$^{-1}$. On the other hand, no such effects are present in the case of Fig.\ \ref{fig:LSI_SED1}(b) because both the shown model spectrum is produced by electrons with $\gamma\ga 2\times 10^3$ as well as the contribution of low-energy electrons to heating is negligible due to the hard injection.

At $\epsilon\ga 10$ GeV, the shape of the photon spectrum starts to be modified by photon-photon \ee\ pair production. The soft photons responsible for the pair opacity originate from both the Be star and its circumstellar disc. However, such discs are transparent in the optical range (since no partial eclipses of Be stars by discs are observed), and thus their emission has to be much weaker than the blackbody spectrum at the disc temperature ($\simeq 0.6T_*$ following \citealt{cb06}). Indeed, detailed calculations of Be disc emission show a substantial excess of emission above that of the star only at $\epsilon\la 0.5$ eV ($\simeq 2\mu$m), see, e.g., \citet*{smj09}. This also roughly agrees with our calculations in Section \ref{free-free}, which show that the disc at $D_0$ becomes optically thin to free-free absorption at $\nu\simeq 3\times 10^{13}$ Hz ($\simeq 0.1$ eV), i.e., the emission is blackbody only up to this energy, and becomes quickly much weaker at higher energies. Thus, we approximated the disc emission as Rayleigh-Jeans at $\leq 0.1$ eV and as optically-thin bremsstrahlung (neglecting the free-bound contribution to emission) at $>0.1$ eV, with the disc truncated at $6R_*$. Though photons at 1 TeV have the threshold for the \ee\ pair production at $\simeq 0.3$ eV, the stellar blackbody spectrum increases fast with energy and the pair opacity is dominated by photons at higher energies. Thus, absorption by disc emission has a relatively minor effect on the shape of the observed TeV emission, except above a few TeV. We notice that our approach differs from that of \citet{or07}, who assumed the disc emits only slightly diluted blackbody at the disc temperature. This disagrees with the Be disc physics, and would strongly overestimate the \gr\ absorption by its emission. Then, R07 have calculated the detailed disc emission in their model, and took it into account for pair absorption. However, they do not present details of their results, in particular, the fractional contribution of this effect to the total absorption.

We take into account the secondary e$^\pm$ pairs produced in \gr\ absorption by adding them to the electron injection rate. However, our calculations underestimate the contribution of the secondary pairs because we consider only the absorption of the radially escaping $\gamma$-rays, for which the optical depth with respect to the pair production is the lowest. This approach is dictated to the one-dimensional nature of our numerical calculations. To take this process more accurately into account, three-dimensional calculations would be required, which is beyond the scope of this work. 

\subsection{Energy losses of $\mathbf{\sim}$10-MeV electrons}
\label{losses}

In the spectral model above, we have identified the major radiative processes taking place in the source. Here, we provide estimates of their relative importance in various energy, density, and spatial regimes. We concentrate on the MeV electron energies, important in the radio and X-ray regimes, and on the model with the Compton emission dominating the 1 keV--10 GeV emission, and the synchrotron process producing the observed radio emission (Fig.\ \ref{fig:LSI_SED1}a). 

Owing to the very high luminosity of the Be star, a substantial part of the X-ray emission can be from Compton scattering of the stellar radiation. At photon energies of $\epsilon \sim\! (1$--10) keV, the electrons have
\begin{equation}
\label{eic}
E\simeq \left(\frac{\epsilon}{3 k T_*}\right)^{1/2} m_{\rm e} c^2\simeq
10\left(\frac{3\times 10^4\,\mbox{K}}{T_*}\right)^{1/2} 
\left(\frac{\epsilon}{3\,\mbox{keV}}\right)^{1/2}\mbox{MeV}.
\end{equation}
Electrons of about the same energy, 
\begin{equation} 
\label{esynch}
E=\frac{(2\upi \nu)^{1/2} m_{\rm e}^{3/2} c^{5/2}}{e^{1/2}B^{1/2}}\simeq
10\left(\frac{1\,\mbox{G}}{B}\right)^{1/2}\left( \frac{\nu}{1\,\mbox{GHz}}\right)^{1/2}\mbox{MeV}, 
\end{equation} 
can producte GHz radio photons at $B\sim 1$ G. 

The high-energy electrons can be retained in the stellar-wind inhomogeneities, see Section \ref{s:short}, by a strong enough magnetic field. Assuming that the electrons diffuse in disordered magnetic field, we find they are retained by the magnetic field inhomogeneities (in the Bohm diffusion regime) for the time, 
\begin{equation}
t_{\rm diff}=\frac{3e B R_{\rm w}^2}{2E c}\simeq 1.5\times 10^7\left(\frac{R_{\rm w}}{10^{11}\,\mbox{cm}}\right)^2
 \left(\frac{B}{1\,\mbox{G}}\right)\left(\frac{10\,\mbox{MeV}}{E}\right) \mbox{s},
\label{t_diff}
\end{equation}
which is much longer than the wind $t_{\rm esc}$, equation (\ref{tesc}). 

Comparing $t_{\rm diff}$ and $t_{\rm esc}$ with the Thomson-regime IC energy loss time of 10~MeV electrons,
\begin{eqnarray}
t_{\rm IC}&=&\frac{3\upi m_{\rm e}^2 c^4 D^2}{\sigma_{\rm T} L_* E}\nonumber\\ &\simeq&
5\times 10^4\left(\frac{10^{38}\,\mbox{erg\,s}^{-1} }{L_*}\right)
\left(\frac{D}{3\times 10^{12}\,\mbox{cm}}\right)^2
\left(\frac{10\,\mbox{MeV}}{E}\right)\,\mbox{s},
\label{t_IC}
\end{eqnarray}
we see that electrons captured by the stellar wind inhomogeneities lose a substantial fraction of their energy before they escape from within the orbit. The synchrotron energy loss time is given by,
\begin{eqnarray}
\label{tsynch}
t_{\rm S}&=&\frac{6\pi m_{\rm e}^2 c^3}{\sigma_{\rm T} 
B^2 E}\simeq 4\times 10^7 \left(\frac{1\,\mbox{G}}{B}\right)^2
\left(\frac{10\,\mbox{MeV}}{E}\right)\,\mbox{s},\\
\label{tsynch2}
&=& {9 m_{\rm e}^{5/2} c^{9/2}\over 4 (2\upi\nu)^{1/2} e^{7/2} B^{3/2} }\simeq 4\times 10^7 \left(\frac{1\,\mbox{G}}{B}\right)^{3/2}
\left(\frac{1\,\mbox{GHz}}{\nu}\right)^{1/2}\mbox{s},
\end{eqnarray}
where equation (\ref{esynch}) has been used in equation (\ref{tsynch2}). The Thomson-regime IC losses dominate over the synchrotron losses up to the ($E$-independent) distance of
\begin{equation}
D_{\rm IC=S}\simeq 8\times 10^{13} \left(\frac{L_* }{10^{38}\,\mbox{erg\,s}^{-1} }\right)^{1/2} \left(\frac{1\,\mbox{G}}{B}\right)\, {\rm cm}.
\label{IC_syn}
\end{equation}
However, this estimate is given for a fixed $B$. If $B=B_0 (D_0/ D)$, as characteristic to stellar winds, and which was assumed in Section \ref{s:spectra}, the IC losses in the Thomson regime dominate everywhere (unless $B_0$ is of the order of tens of G). 

At the density of the stellar wind, see equations (\ref{n_disc}), (\ref{n_wind2}), the electrons may suffer from strong Coulomb energy losses. The Coulomb loss time is given by,
\begin{eqnarray}
\label{eq:tcoul}
t_{\rm C} &\simeq& \frac{4E}{3\sigma_{\rm T} m_{\rm e} c^3 n_{\rm e} }\ln^{-1} \frac{1.24 \upi E m_{\rm e}^2 c^2}{h^2 e^2 n_{\rm e}}\nonumber\\
&\simeq& 8\times 10^4 \left(\frac{3\times 10^{8}\,\mbox{cm}^{-3}}{n_{\rm e}}\right) \left(\frac{E}{10\,\mbox{MeV}}\right)\,\mbox{s}
\label{t_Coul}
\end{eqnarray}
\citep{gould75}, where $n_{\rm e}=n_{\rm i}\mu_{\rm i}/\mu_{\rm e}$. The Coulomb loss time becomes shorter than $t_{\rm IC}$, equation (\ref{t_IC}), below
\begin{equation}
E_{\rm C}\simeq 8 \left(\frac{D}{3\times 10^{12}\,\mbox{cm}}\right)
\left(\frac{n_{\rm e}}{3\times 10^{8}\,\mbox{cm}^{-3}}\right)^{1/2}\mbox{MeV.}
\label{E_Coul}
\end{equation}
Most (but not all) of the power injected in the form of electrons with $E\la E_{\rm C}$ goes into heating of the stellar wind (see Section \ref{wind}) rather than into the synchrotron and IC emission. Equating the Coulomb break energy to the energy of electrons emitting X-ray IC radiation at $\epsilon$, equation (\ref{esynch}), we find the critical density,
\begin{equation}
n_{\rm IC=C}\simeq 4\times 10^{8} \left(\frac{\epsilon}{3\,\mbox{keV}}\right)
\left(\frac{D}{3\times 10^{12}\, \mbox{cm}}\right)^{-2} 
\mbox{cm}^{-3}.
\label{n_Coul}
\end{equation}
At higher densities, the X-ray spectrum is hardened by the effect of Coulomb losses on the electron distribution. This effect is only moderate in the isotropic wind, see equation (\ref{n_wind2}). However, Coulomb losses will strongly dominate in the dense disc, see equation (\ref{n_disc}). As both $n_{\rm IC=C}$ and $n_{\rm w}$ scale as $D^{-2}$, the ratio between the IC and Coulomb loss times is independent of the distance,
\begin{equation}
{t_{\rm IC}\over t_{\rm C}} \simeq 0.7\left(\frac{\dot M_{\rm w}}{ 10^{-8}\,\msun\, \mbox{yr}^{-1}}\right) \left(f\over 0.1\right)^{-1}
\left(\frac{v_\infty}{10^8\, \mbox{cm\,s}^{-1}}\right)^{-1}
\left(\frac{\epsilon}{3\,\mbox{keV}}\right)^{-1}.
\label{IC_Coul}
\end{equation}

Then, the density above which Coulomb losses dominate over synchrotron losses can be found from equations (\ref{esynch}), (\ref{tsynch2}) and (\ref{t_Coul}),
\begin{equation}
n_{\rm S=C}\simeq 5\times 10^{5} \left(\frac{\nu}{1\,\mbox{GHz}}\right)
\left(B\over 1\,{\rm G}\right)^2 
\mbox{cm}^{-3}.
\label{n_Coul_syn}
\end{equation}
For the fast isotropic wind, equation (\ref{n_wind2}), this density is achieved at a distance,
\begin{eqnarray}
D_{\rm S=C}&\simeq& 7\times 10^{13}\left(\frac{B}{1\,\mbox{G}}\right)^{-1} \left(\frac{\dot M_{\rm w}}{ 10^{-8}\,\msun\, \mbox{yr}^{-1}}\right)^{1/2}
\nonumber\\
&\times&  \left(f\over 0.1\right)^{-1/2}
\left(\frac{v_\infty}{10^8\, \mbox{cm\,s}^{-1}}\right)^{-1/2}
\left(\frac{\nu}{1\,\mbox{GHz}}\right)^{-1/2}
\mbox{cm}.
\label{D_Coul_syn}
\end{eqnarray}  
Again, this estimate is given for a fixed $B$. If $B=B_0 (D_0/ D)$, the Coulomb losses dominate over synchrotron ones everywhere (for radio-emitting electrons and unless $B_0$ is of the order of tens of G).

\section{Structure of the compact pulsar wind nebula}
\label{structure}

VLBA monitoring over an entire orbital cycle (DMR06) have
revealed an extended radio source of a variable morphology with the
overall size of $\sim$5--10 mas\footnote{On fig.\ 3 of DMR06, 1 cm corresponds to 6 mas, V. Dhawan, personal communication.} at 8.3 GHz ($\epsilon \simeq 3\times 10^{-5}$ eV], which can be identified with the compact PWN of \lsi, similar to the compact PWN of PSR B1259--63 \citep{neronov07}. At $d=2$ kpc, the source size corresponds to $D\simeq (1.5$--$3)\times 10^{14}$ cm. However, note that the dynamic range of the images in DMR06 is up to a factor of 64, i.e., the flux at the boundary of a radio image may be only a $\sim$2 per cent of that at its centre. Thus, an e-folding distance is $\sim$4 times less than the
above overall size, i.e., $D_{\rm PWN}\sim\! (4$--$8)\times 10^{13}$
cm. This is in good agreement with our theoretical modelling of the radio flux profiles in Section \ref{s:spectra}, see Fig.\ \ref{fig:brightness}(a).

For $v_{\rm w}\sim 10^8$ cm s$^{-1}$, the wind escape time scale,
$D_{\rm PWN}/v_{\rm w}\sim 5\times 10^5$ s. Thus, it is substantially
shorter than the orbital period, $P\simeq 2.3\times 10^6$ s. The
strong variability of the radio morphology of the source over the
course of an orbit can be then associated with this time scale. 

An interesting issue is here what is the cause of the gradient of the radio source structure on the scale of $D_{\rm PWN}$. It is possible that the decline of the flux with distance is due to the energy loss of electrons in the magnetic field there. Equating the synchrotron energy loss time, equation (\ref{tsynch2}), to $D_{\rm PWN}/v_{\rm w}$, we obtain a relatively strong magnetic field, $B\sim 10$ G. On the other hand, this assumption does not need to hold, the magnetic field can be much weaker, and the e-folding scale may correspond to a gradient of the magnetic field, as we have assumed in Section \ref{s:spectra}. Consequently, the radio-emitting electrons lose most of the energy within the radio source and the radio radiative efficiency is high in the former case, or they lose only a small fraction of their energy and the efficiency is low in the latter case. 

Another important piece of information is the position of the peak or
centroid of the radio emission (DMR06). At 8.3 GHz, the
distance of the emission peak from the Be star varies with the orbital
phase in the range of $D\simeq (0.5$--$8)\times 10^{13}$ cm, and it is
always outside the binary orbit. At 2.3 GHz, the centroid distance
equals $D\simeq (1$--$10)\times 10^{13}$ cm. This roughly agrees with our first model of Section \ref{s:spectra}. There is also a significant lag of the 2.3 GHz emission with respect to the 8.3 GHz one. Similarly to the case of the flux spatial gradient, the lag can be due to the electron energy loss if the magnetic field is high, or due to the gradient of the $B$ field if it is low. 

The avoidance of the orbital region by the radio emission also approximately agrees with our estimates of equations (\ref{tau_disc_ff}--\ref{D_ff}) of that region being optically thick to free-free absorption by the stellar wind. An important unknown factor, however, is the degree of the heating of the stellar wind by the X-rays and $\gamma$-rays, as discussed in Section \ref{wind}. 

The alternative model of accretion/jet for the evolving radio structure has recently been discussed in some detail by \citet{mk09}. It can explain some aspects of the evolution of the orbital flux periodicity, but the rotation of the structure with the orbital phase remains unexplained.

\section{The $\boldsymbol{\gamma}$-ray light curve}
\label{gamma}

The IC cooling time is the shortest at the boundary between the Thomson and Klein-Nishina regimes for Compton scattering, $3kT_* E \sim (m_{\rm e}c^2)^2$, when $E\simeq 30$ GeV and $t_{\rm IC}\simeq 10 (D/3\times 10^{12}\,\mbox{cm})^2$ s, see equation (\ref{t_IC}). The IC
scattering on TeV electrons proceeds in the Klein-Nishina regime. In
this regime, $\epsilon\simeq E$, and the electron energy loss time
grows with energy \citep{bg70},
\begin{eqnarray}
\label{tkn}
t_{\rm KN} &\simeq& {2 E D^2 h^3\over \sigma_{\rm T}\upi^3(m_{\rm e} c k T_* R_*)^2 }\ln^{-1}{0.552 E k T_*\over m_{\rm e}^2 c^4}\nonumber\\
&\simeq& 10^2 \left(\frac{E}{1\,\mbox{TeV}}\right)
\left(\frac{D}{3\times 10^{12}\,\mbox{cm}}\right)^2\mbox{s,}
\end{eqnarray}
where the density of the blackbody photons has been diluted by the
$(D/R_*)^2$ factor. This time scale is much shorter than the energy loss
time for 10~MeV electrons, equation (\ref{t_IC}). The highest energy electrons are not retained by the stellar wind clumps because their diffusion time is very short, equation (\ref{t_diff}). Thus, the electrons emitting at the highest energies are initially present only in the vicinity of the initial injection region, around the bow-shaped average contact surface of the pulsar and stellar winds. Thus, the highest-energy emission from the pulsar/stellar wind interaction region can be modelled, in the first approximation, by a regular bow-shaped contact surface.

\begin{figure} 
\centerline{\includegraphics[width=\linewidth]{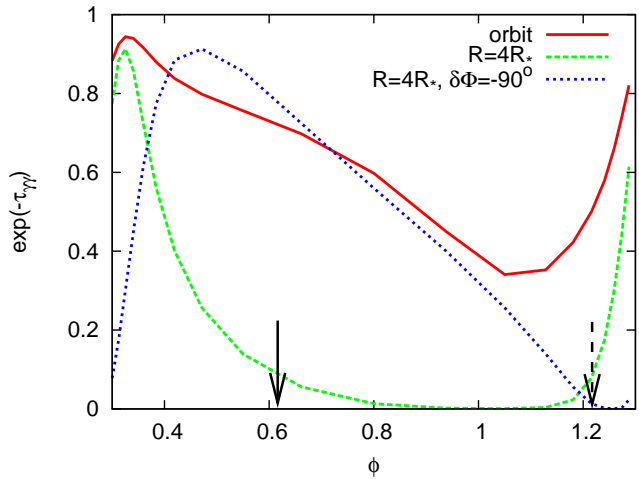}}
\caption{Attenuation of 1-TeV \gr s produced in interaction of the pulsar wind with the disc of Be star as a function of the orbital phase. The red solid curve is for photons emitted isotropically at the position of the pulsar. The blue dotted curve is for isotropic emission at $D=4R_*$ (close to the boundary of the truncated disc around the Be star along the line connecting the two stars, see the blue box in Fig.\ \ref{fig:scheme}).  The green dashed curve is for a point at $D=4R_*$ but shifted by $\delta\Phi=-90\degr$, see the green box in
Fig.\ \ref{fig:scheme}. The solid arrow shows the approximate observed phase of the maximum of the VHE \gr\ light curve. The dashed arrow shows the phase of the second maximum, which is not observed because of the absorption.
}
\label{tau_lc} 
\end{figure}

Within this model, the \gr\ emitting plasma flows along the bow-shaped contact surface of the pulsar and stellar winds, see Fig.\ \ref{fig:scheme}. The contact surface is asymptotically a cone with a half-opening angle, $\theta_\infty$, which depends on the ratio of the momentum fluxes of the pulsar and stellar winds,
\begin{equation}
\theta_\infty\simeq 180\degr \frac{\eta}{1+\eta}
\end{equation}
where $\eta=\dot E_{\rm pulsar}/(\dot M_{\rm w} v_{\rm w} c)$ (R07).  The shocked pulsar wind can flow along the cone with a large bulk Lorentz factor \citep{bogovalov08}, leading to a beaming of the emission along the surface. Thus, a maximum of the \gr\ emission is expected when a part of the cone passes through the line of sight. The cone can pass through the line of sight twice per orbit, so, in principle, two maxima of the \gr\ light curve are expected. The true anomaly corresponding to those two events is at equal angles from the inferior conjunction, $\Phi=\Phi_{\rm inf}\pm \Delta \Phi$. The MAGIC \citep{albert06} and VERITAS \citep{veritas} observations of the source indicate that the maximum of the TeV light curve is at the phase interval of $0.6< \phi <0.7$, see the solid arrow in Fig.\ \ref{tau_lc}. This interval corresponds to $153\degr < \Phi < 167\degr$. Taking $\Phi_{\rm inf}=33\degr$ (see Section \ref{parameters}) and choosing the plus sign above, we obtain $120\degr < \Delta \Phi < 134\degr$. 

The shift, $\Delta\Phi$, is related to $\theta_\infty$ and the inclination, $i$, via \citep{neronov08}
\begin{equation}
\cos \Delta\Phi={\cos \theta_\infty\over \sin i}.
\label{Delta_Phi}
\end{equation}
We can see that $|\cos \theta_\infty|\leq \sin i$ is required for the cone to pass through the line of sight. For $i=59\degr$ (Section \ref{parameters}) and the above range of $\Delta\Phi$, we obtain $115\degr < \theta_\infty < 127\degr$. Interpreting this measurement of the phase of the $\gamma$-ray maximum as being due to the passage of the cone through the line of sight, we find the ratio of momentum flux in the pulsar wind to that in the stellar
wind of $1.8\la \eta\la 2.4$. The suggested phases of the two maxima are shown 
by the arrows in Fig.\ \ref{tau_lc}.

The above simple geometrical model predicts a second maximum of the TeV light curve, at the phase of the second passage of the emission cone through the line of sight. However, similarly to the case of the X-ray light curve, the second maximum (shown by the dashed arrow in Fig. \ref{tau_lc}) is suppressed because of the compactness of the system. In the case of the TeV light curve, the suppression is due to absorption of  the emitted \gr\ photons in result of the pair production in the photon field of the Be star. 

The effect of the $\gamma\gamma$ pair production on the properties of the TeV emission from the system was previously studied by \citet{bednarek06a,bednarek06b} for the case of emission from a jet (assuming that the compact object is a black hole which emits jet in the direction normal to the orbital plane) and by \citet{dubus06a}, under the assumption that the TeV flux comes directly from the compact object. 

In the case of the interacting pulsar and stellar winds, the region of $\gamma$-ray production is extended. Even in the simple model with a smooth stellar wind, see Section \ref{s:simple}, the \gr\ emission region extends along the bow-shaped interaction surface. Thus, absorption of the \gr\ emission emitted toward an observer varies strongly across the surface. The strongest absorption is for photons emitted around the apex of the bow-shaped surface, closest to the massive star. The situation becomes even more complicated when the stellar wind is clumpy. Thus, we can estimate only limits within which the optical depth, $\tau_{\gamma\gamma}$, varies across the emission region at different orbital phases.

The results of such calculations are shown in Fig.\ \ref{tau_lc}, which takes into account the anisotropy of UV emission from Be star. The red solid curve shows the attenuation of 1-TeV photons emitted at the position of the compact object. This point can be considered as a rough outer boundary of the pulsar/stellar wind interaction region. (Note that the shape of the curve slightly differs from that calculated by \citealt{dubus06a}, because we use the current estimates of the orbital parameters of G07, whereas he used the results of C05.) The blue dotted curve shows the attenuation of photons emitted at the point situated at the distance of $D=4R_*\simeq 2\times 10^{12}$ cm from the centre of the Be star on the line connecting it to the compact object. This distance is a rough estimate of the radius of the truncated Be-star disc, see Section \ref{wind}, and we also used it in our spectral calculations in Section \ref{s:spectra}. Finally, the green dashed curve shows the attenuation of emission produced at $D=4R_*$, but $\delta\Phi =-90\degr$ away from the direction of the line connecting the two stars. These three points at $\phi=0.7$ are shown by the red, blue and green box, respectively, in Fig.\ \ref{fig:scheme}.

We see that the orbital dependence of the attenuation varies by a large factor across the \gr\ emission region. The transmitted fraction reaches the maximum at the inferior conjunction for the emission on the line connecting the two stars, but that maximum is shifted to a later phase for the blue dotted curve. An accurate calculation needs to assume the detailed dependence of the emissivity on both the position and the phase. This cannot be done as the morphology of such extended sources remains so far unknown.

\section{Conclusions}
\label{conclusions}

We have shown that the compact PWN model developed here can explain a number of observational properties of \lsi, from the radio to the TeV \gr\ energy band. The main new feature of this model is to account for the clumpiness of the fast wind component from the Be star, which slows down the escape of the high-energy electrons from the system. Also, due to plasma instabilities, the relativistic electrons are retained in the moving clumps of the stellar wind and carried away with them. The presence of time-dependent X-ray emitting clumps explains the X-ray variability observed on time scales much shorter than the orbital period. On the other hand, our model has not resolved yet the issue of the momentum flux of the pulsar wind being significantly higher than that of the Be wind, which presents a problem for interpretation of the observed radio structures (as pointed out by R07). 

The electrons lose energy in the IC, synchrotron and Coulomb processes. We have presented detailed inhomogeneous spectral models, reproducing the average spectra of the object. The 1 keV--10 GeV spectral range is dominated either by synchrotron or IC emission, depending on the form of the electron injection. Our models reproduce very well, in particular, the sharp spectral cut-off seen in the GeV range by \fermi. 

We have also provided detailed formulae and calculations comparing the relative importance of those processes as functions of the electron energy and the position in the source, measured by the distance from the Be star. Coulomb energy losses are important for MeV electrons. The radio emission is suppressed by strong free-free absorption, up to the distance of about an AU from the source, in agreement with the radio data. The free-free absorption also leads to a strong attenuation of the pulsed radio emission from the neutron star. Compton losses dominate in an inner part of the fast wind. We also find that a part of the wind surrounding the high-energy source is strongly heated by both Compton and Coulomb processes. 

We find the most likely mechanism explaining the observed orbital modulation at TeV energies is anisotropy of emission, with relativistic electrons accelerated along the surface at which the pressure of the pulsar wind equals the average pressure of the stellar wind. In this model, pair absorption of the TeV emission plays a role by suppressing the one of the two expected maxima. 
 
\section{Acknowledgments}

The authors thank G. Dubus, J. Miko{\l}ajewska, K. St{\c e}pie{\'n} and A. Szostek for valuable discussions, and the referee for valuable comments and suggestions. AAZ has been supported in part by the Polish MNiSW grants NN203065933 and 362/1/N-INTEGRAL/2008/09/0.

\label{lastpage}

\end{document}